\documentclass[%
 preprint,
  aps,
  prd,
 eqsecnum,
amsmath,
amssymb,
showpacs
]{revtex4-1}
\usepackage[english]{babel}
\usepackage{hyperref}

\begin{document}
  \title{Weak field limit in vierbein-Einstein-Palatini formalism and Fierz-Pauli Equation}
  \author{Subhasish Chakrabarty}
  \email{subhasish.chy@bose.res.in}
  \affiliation{S. N. Bose National Centre for Basic Sciences\\
  Block - JD, Sector - III, Salt Lake, Kolkata - 700098}
  \author{Amitabha Lahiri}
  \email{amitabha@bose.res.in}
  \affiliation{S. N. Bose National Centre for Basic Sciences\\
  Block - JD, Sector - III, Salt Lake, Kolkata - 700098}

\begin{abstract}
We consider the weak field limit of gravity in the vierbein-Einstein-Palatini 
formalism, find the action and the equations for perturbations around an 
arbitrary background, and compare them with the usual metric perturbation 
equations. We also write the Fierz-Pauli equations for massive gravitons
on an arbitrary curved background in this formalism.
\end{abstract}


\maketitle 

%%%%%%%%%%%%%%%%%%%%%%%%%%%%%%
\section{Introduction}
%%%%%%%%%%%%%%%%%%%%%%%%%%%%%
General relativity is a non-linear theory.
A linearized approximation to the theory can be obtained in
by considering the gravitational field to be so weak that 
the space-time metric can be considered as a background 
metric plus a perturbation~\cite{Carroll:2004st}. 
Einstein's equation turns out to be a linear second order 
equation in the perturbation. 

Another formulation of general relativity uses tetrads and spin connections
instead of metric and Christoffel symbols. In this formulation,
 an internal Minkowski space is attached to each point of
space-time, isomorphic to the tangent space of the space-time
manifold at that point. Linear isomorphisms between the
tangent space and the internal space are given by tetrads, also known 
as vierbeins. Any connection on this bundle, called a spin connection,
gives the parallel transport of the sections of the internal space. 
The curvature of this connection is related to the space-time 
curvature, which allows us to write the Einstein-Hilbert action 
of general relativity in terms of the tetrads and the spin connection.
This formulation is known as vierbein-Einstein-Palatani (VEP) 
formalism~\cite{Baez:1995sj,Peldan:1993hi,Hehl:1994ue}. 
In this paper we find the weak field limit of the VEP 
action and equations. 

The idea of perturbing the vierbein
is not a new one (a representative list is~\cite{Bleyer:1982,
Nibbelink:2006sz, Zheng:2010am, Chen:2010va, Dzhunushaliev:2014daa, 
Wu:2012hs, Hinterbichler:2012cn, Alexandrov:2013rxa}), 
but our paper differs from earlier work 
in an important manner. In all papers dealing with vierbein 
perturbations that we have been able to find, the perturbations
were considered around a flat background spacetime. 
The background spacetime and the internal Lorentz space are then 
identical, and as a result
the background tetrad can be written as a Kronecker delta\,, 
$ e^I_\mu = \delta^I_\mu $. Furthermore, the background connection is 
then trivially flat and torsion-free, which means no attention is 
paid to the independent nature of the spin connection. This paper
is different from earlier work in both ways -- we assume a general curved
background and an independent spin connection, so tetrad fields are
not constant, while the absence of torsion is implemented by the equations
of motion for the spin connection only in the absence of fermionic matter.
When fermions are coupled to the spin connection, torsion remains
as a part of the spin connection, and the latter cannot be completely
eliminated from the perturbation equations.  As far as we are aware, 
this is the first work to write down tetrad perturbation equations 
on an arbitrary curved background, and in presence of torsion. While gravity 
with torsion has a long history (see~\cite{Gogala:1980,
	Hayashi:1979qx, 
Deffayet:2011uk, Nair:2008yh, Nikiforova:2009qr} for some recent work 
which are close to our work in spirit), we find perturbation equations for 
tetrads and spin connection, and write the torsion 
in terms of these variables.

Another motivation for looking at vierbein perturbations is 
to investigate massive gravity~\cite{Hinterbichler:2011tt, Babichev:2009jt,
Alberte:2010qb, Deffayet:2012nr, Bernard:2015mkk, Bernard:2014bfa, 
Deser:2015wta} 
in this formalism. 
%%% A theory of massive gravity is one that describes propagating 
%%% massive spin-2 particles and is the modification of gravity 
%%% in the infrared limit. 
Supernova data~\cite{Perlmutter:1998np, 
Riess:1998cb} reveal that the universe is in a phase of accelerated 
expansion, which can be explained by positing a dark energy component 
along with matter and dark matter in cosmological models based on
general relativity. Dark energy is most simply modeled by the 
addition of a constant term, called the cosmological constant, 
in general relativity. However, there is no convincing model for
the cosmological constant itself, which looks like a vacuum energy
density, but the value of the cosmological constant differs from 
the value obtained from the vacuum energy in quantum field 
theory~\cite{Weinberg:1988cp} by a factor 
of$~\sim 10^{65} $. An alternative suggestion is that general 
relativity itself is modified in the infrared which may remove 
the need for dark energy to explain the acceleration. One particular 
infrared modification of general relativity is one with massive
gravitons, i.e. one where metric perturbations obey a massive
wave equation. 

The story of massive gravity goes back a long way.
Fierz and Pauli~\cite{Fierz:1939zz, Fierz:1939ix} proposed an action
that describes a free massive graviton on a flat background. 
The Fierz-Pauli equations 
are linear in the metric perturbation, but it was shown later that 
the massless equation cannot be reached by a continuous limiting 
procedure~\cite{vanDam:1970vg, Zakharov:1970cc}. The problem seemed to lie in
the linearization procedure~\cite{Vainshtein:1972sx}, but it was soon 
shown that any non-linear version introduces ghost 
instability~\cite{Boulware:1973my}. Since then there have been several 
attempts to find a consistent theory of massive gravity 
and significant progress has been made recently, using a
bimetric formulation~\cite{deRham:2010ik, 	deRham:2010kj, 
Hassan:2011vm, 	Hassan:2011hr, Hassan:2011zd, Nomura:2012xr, 
Baccetti:2012re, Berg:2012kn, Baccetti:2012bk, Martin-Moruno:2013wea, 
Paulos:2012xe}. 
An important building block 
of these constructions is a matrix of the form $\sqrt{g^{-1} f}\,,$
where $g$ is the metric and $f$ is the auxiliary metric, and the square
root is defined so that $\sqrt{g^{-1} f}\, \sqrt{g^{-1} f} = g^{-1} f\,.$
It is this `square root of metric' that hints at a possible reformulation 
in terms of tetrads, since the metric is related to the 
tetrads by $g_{\mu \nu} = e^I_\mu e^J_\nu  \eta_{IJ}\,.$ 
The tetrad formulations of bimetric gravity that are available in the 
literature has two sets of tetrads, one of which acts as an auxiliary
tetrad and the other is the dynamical one. As in the case of tetrad
perturbations, the auxiliary tetrad is usually taken to be the
Kronecker delta, and the torsion is ignored. One may expect that
in the bi-tetrad formulation of the full theory of massive gravity 
the auxiliary tetrad will correspond to an arbitrary background, 
including torsion.

%%% Furthermore, since the tetrad formulation of gravity closely 
%%% resembles gauge theories~\cite{Hehl:1994ue}, it may be possible or indeed
%%% convenient to express massive gravity as a massive gauge theory in terms 
%%% of tetrads and spin connection.

This paper is, however, somewhat less ambitious. Here we consider
classical perturbations of gravity in the VEP formalism, and then
construct the Fierz-Pauli equation, and the corresponding action, 
in terms of these perturbations. 
%%% We also consider the first order action only (unlike \cite{Nair:2008yh}), 
%%% which is sufficient upto the predictions of Einstein's GR. 
It should be mentioned that while the vierbein formulation of bimetric 
massive gravity has also received some attention recently~\cite{Hinterbichler:2012cn,
	Alexandrov:2013rxa, Nibbelink:2006sz}, our approach cannot be directly compared 
to these theories, again because the mass term in this paper is written 
for perturbations around an arbitrary curved background. We note that
bimetric gravity for an arbitrary background metric has been investigated
recently~\cite{Bernard:2014bfa, Bernard:2015mkk}, although not in 
the vierbein formalism, and perturbations were not considered. It may 
be expected that linearization of that formulation, in terms of tetrads,
can be related to what we find below. We leave a careful 
comparison, as well as an investigation into ghost degrees of
freedom in this formalism, for future work.
 
Let us briefly recapitulate the formalism of metric perturbation 
theory. In this, the space-time metric is written as a
background metric plus a perturbation. Often we are interested in a
flat background, in which case 
\begin{equation}\label{metric}
g_{\mu\nu} = \eta_{\mu\nu} + h_{\mu\nu}\,,
\end{equation}
with $|h_{\mu\nu}|\ll 1\,$ for each component. The inverse is given by 
\begin{equation}
g^{\mu\nu} = \eta^{\mu\nu} - h^{\mu\nu}\,,
\end{equation}
where all indices are raised and lowered with $\eta\,.$

The linearized Riemann and Ricci tensors, and the Ricci scalar, are thus
\begin{align}
R_{\mu\nu\rho\sigma} &= \partial_\rho\partial_\nu h_{\mu\sigma}
+ \partial_\sigma\partial_\mu h_{\nu\rho}
- \partial_\sigma\partial_\nu h_{\mu\rho}
- \partial_\rho\partial_\mu h_{\nu\sigma}\,,\\
%\end{equation}
%
%\begin{equation}
R_{\mu\nu} &= \frac{1}{2}\left[\partial_\sigma\partial_\nu
h^\sigma_{~\mu} + \partial_\sigma\partial_\mu h^\sigma_{~\nu}
- \partial_\mu\partial_\nu h - \square h_{\mu\nu}\right]\,, \\
%\end{equation}
%
%\begin{equation}
R &= \partial_\rho \partial_\sigma h^{\rho\sigma} - \square h\,,
\end{align}
where $h$ is the trace of the perturbation, $h = \eta^{\mu\nu}h_{\mu\nu}\,.$
The linearized Einstein equation in vacuum follows from these,  
\begin{align}\label{fierz-pauli}
\frac{1}{2}[ \partial_\sigma\partial_\nu h^\sigma_{~\mu}
+ \partial_\sigma\partial_\mu h^\sigma_{~\nu}
- \partial_\mu\partial_\nu h - \square h_{\mu\nu} 
 - \eta_{\mu\nu}\partial_\rho \partial_\sigma h^{\rho\sigma}
 +\eta_{\mu\nu} \square h ] = 8\pi G T_{\mu\nu}\,,
\end{align}
which is the linearized form of $ G_{\mu\nu} = 8\pi G T_{\mu\nu}$. 
Usually we are concerned with the vacuum equation, $ G_{\mu\nu} = 0 $\,.
The linearized version written above is for a special situation 
where we have taken the background to be flat. We will consider 
perturbations around a general curved background. Then 
let us denote the background metric by $ \bar{g}_{\mu\nu} $ 
so that the total metric is written as
$ g_{\mu\nu} = \bar{g}_{\mu\nu} + h_{\mu\nu} $. Then the 
Christoffel symbol is 
\begin{equation}
\Gamma^\alpha_{\mu\nu} =\bar{\Gamma}^\alpha_{\mu\nu} 
- \frac{1}{2}h^{\alpha\lambda}\left[\partial_\mu \bar{g}_{\lambda \nu} 
+ \partial_\nu \bar{g}_{\mu \lambda} - \partial_\lambda 
\bar{g}_{\mu\nu}\right] + \frac{1}{2}\bar{g}^{\alpha\lambda}
\left[\partial_\mu h_{\lambda \nu} + \partial_\nu h_{\mu \lambda} 
- \partial_\lambda h_{\mu\nu}\right] 
%\notag\\&
+\mathcal{O}(h^2),
\end{equation}
where now the indices are lowered and raised by $\bar g_{\mu\nu}$
and its inverse $\bar g^{\mu\nu}$\,, respectively, and quantities
pertaining to the background are denoted by a bar. 

By calculating the Einstein tensor using these Christoffel symbols, we 
can write the Einstein equations for an arbitrary background,
\begin{equation}\label{metric_pert_matter}
\bar{R}_{\mu\nu} - \frac{1}{2} \left(\bar{g}_{\mu\nu}
\bar{g}^{\alpha\beta} + h_{\mu\nu}\bar{g}^{\alpha\beta}
-\bar{g}_{\mu\nu} h^{\alpha\beta}\right)\bar{R}_{\alpha\beta}
%& \notag \\
+ R^1_{\mu\nu} - \frac{1}{2}\bar{g}_{\mu\nu}
\bar{g}^{\alpha\beta}R^1_{\alpha\beta} = 8\pi G T_{\mu\nu}\,,
\end{equation}
where the Ricci tensor $\bar R_{\mu\nu}$ 
is derived from the background metric $\bar g_{\mu\nu}\,,$ 
and the quantity $R^1_{\mu\nu}$ is the part of the Ricci 
tensor linear in $h\,.$ This equation, however, contains the 
background equation $ \bar{R}_{\mu\nu} - \frac{1}{2} \bar{g}_{\mu \nu} 
\bar{R} = 8 \pi G \bar{T}_{\mu\nu} $ which needs to be eliminated. 
Note that if the stress-energy tensor is calculated from a matter action
by varying the metric, $T_{\mu\nu}$ and $ \bar{T}_{\mu\nu}$ are
not equal in general. Thus we can finally write the equation for gravitational
perturbations as
\begin{align}\label{metric_pert_matter.2}
\frac{1}{2} \left( \bar{g}_{\mu\nu} h^{\alpha\beta} - 
h_{\mu\nu}\bar{g}^{\alpha\beta} \right)\bar{R}_{\alpha\beta}
%& \notag \\
+ R^1_{\mu\nu} - \frac{1}{2}\bar{g}_{\mu\nu}
\bar{g}^{\alpha\beta}R^1_{\alpha\beta} &= 8\pi G 
\left( T_{\mu\nu} - \bar{T}_{\mu\nu}\right)\,.
\end{align}
In this paper we will consider the VEP formulation of Einstein gravity, 
and find the tetrad equivalent of the linearized Einstein's equation. We will 
consider perturbations on a general curved background, and treat
the perturbations of tetrads and spin connection independently. 
When the background spacetime is flat, we recover the standard
tetrad perturbation equations. The paper is organized in the following way.
In Sec.~\ref{VEP} we briefly recall the 
VEP formulation of gravity, to fix the notation and conventions. 
In Sec.~\ref{pert} we find the equations for small perturbations 
of the vierbein around an arbitrary background space-time with matter. 
In Sec.~\ref{fields} we consider different fields as matter source. 
In particular, we find that for the real scalar field and the 
electromagneic field the perturbation equations can be written
purely in terms of the vierbein perturbations. For a fermionic 
field on the other hand, there is a torsion component to the spin
connection, so the latter cannot be eliminated from the perturbation
equations. Finally in Sec.~\ref{mass} we write a mass term for the 
vierbein perturbations, leading to the Fierz-Pauli equation and 
its generalization for a curved background. 

%%%%%%%%%%%%%%%VEP%%%%%%%%%%%%%%%%%%%%%%%
\section{Vierbein-Einstein-Palatini formalism}\label{VEP} 
 %%%%%%%%%%%%%%%%%%%%%%%%%%%%%%%%%%%%%%%%%
In the vierbein-Einstein-Palatini formalism, the variables
for gravity are the vierbein or tetrads $e^I_\mu\,,$ and the spin
connection $A^{IJ}_\mu$. We will denote space-time indices by lowercase
Greek letters and internal indices by uppercase Roman 
letters. The internal space is a 4-dimensional flat space with 
metric $\eta_{IJ} = (-1, 1, 1, 1) $ attached
to each point of space-time. Raising and lowering of the internal 
indices are done by $ \eta \,,$ while space-time indices are raised 
and lowered by the space-time metric $g\,,$ which is also of signature 
$(-+++)\,.$
The tetrads are defined to be orthonormal, 
\begin{equation}
g^{\mu \nu} e^I_\mu e^J_\nu = \eta^{IJ},
\end{equation}
which we can rewrite as
\begin{equation}
e^I_\mu e^\mu_J=\delta^I_J\,, \qquad
%%% \end{equation}
%%% and
%%% \begin{equation}
e^\mu_I e^I_\nu=\delta^\mu_\nu,
\end{equation}
identifying $ e^\mu_I \equiv  \eta_{IJ} g^{\mu\nu} e^J_\nu\,$ as the
co-tetrad. It is easy to see that the determinants are related by 
$ |e| = \sqrt{-g} $.

A connection $ D $ on this bundle is defined by its action on 
any smooth section $ S $,  
\begin{equation}
D_\mu S^I=\partial_\mu S^I + A^I_{\mu J} S^J\,,
\end{equation}
where $ A^I_{\mu J}$ are the components of what is called the spin connection.
It follows from definition that $A^{IJ}_\mu$ is antisymmetric,
\begin{align}\label{a1.3}
\nonumber
0 = D_\mu \eta^{IJ} &=  \partial_\mu \eta^{IJ}-A^I_{\mu K}
\eta^{KJ}-A^J_{\mu K}  \eta^{IK} \notag \\ 
\Rightarrow \qquad\qquad  A^{IJ}_\mu &= -A^{JI}_\mu.
\end{align}
The curvature of $ D $ can be written as 
\begin{align}
F^{IJ}_{\mu\nu} &=  \left[ D_\mu\,, D_\nu\right]^{IJ} \notag \\
& = \partial_\mu A^{IJ}_\nu- \partial_\nu
A^{IJ}_\mu +  A^I_{\mu K}A^{KJ}_\nu-A^I_{\nu K} A^{KJ}_\nu
\notag \\
& =  \partial_\mu A^{IJ}_\nu- \partial_\nu A^{IJ}_\mu +
[A_\mu,A_\nu]^{IJ}\,.
\end{align}
In order to write the connection on space-time,  we define a set of 
Christoffel symbols
\begin{equation}
\Gamma^\alpha_{\mu \nu}= e^\alpha_I \partial_\mu e^I_\nu +
A^I_{\mu J} e^J_\nu e^\alpha_I\,. 
\label{gammadef}
\end{equation}
This leads to a metric-compatible connection, as we see from the
following calculation, 
\begin{align}
\nonumber
\nabla_\alpha g_{\mu \nu}&=\partial_\alpha g_{\mu\nu}
-\Gamma^\beta_{\alpha \mu} g_{\beta\nu}
-\Gamma^\beta_{\alpha \nu} g_{\mu \beta}\\ 
\nonumber
&= \eta_{_{IJ}} \partial_\alpha (e^I_\mu e^J_\nu)
- \eta_{IJ} e^J_\nu \partial_\alpha e^I_\mu 
-  \eta_{_{IJ}} e^I_\mu\partial_\alpha e^J_\nu 
%\\ \nonumber&\qquad 
-A^I_{\alpha J}  \eta_{_{IL}} e^J_\mu
e^L_\nu 
- A^I_{\alpha J} \eta_{_{KI}} e^J_\nu   e^K_\mu \\ 
%\nonumber
&= -A^{IJ}_\alpha e_{J \mu} e_{I \nu}
%\\ \nonumber&~
-A^{IJ}_\alpha e_{J \nu} e_{I \mu} %\\
= 0\,,\label{metricity}
%\eta_{_{IJ}} \partial_\alpha (e^I_\mu e^J_\nu)
\end{align}
where we have used the antisymmetry of the spin-connection $A$ in 
the last step. 

We can now calculate the  Riemann tensor in terms of the
tetrads and the spin connection,
\begin{equation}
R^\rho_{~\sigma \mu \nu}=F^I_{\mu \nu J} e^\rho_I e^J_\sigma\,,
\end{equation}
thereby getting the Ricci tensor and Ricci scalar respectively as
\begin{align}
R_{\sigma \nu} &= F^I_{\mu \nu J} e^\mu_I e^J_\sigma\,,\label{rictensor}\\
R &= F^{IJ}_{\mu \nu} e^\mu_I e^\nu_J\,.\label{ricscalar}
\end{align}
In the Einstein-Hilbert action for gravity, we replace the Ricci scalar by 
Eq.~(\ref{ricscalar}), and the metric determinant by that of tetrads,  to write 
the action as
%%% The VEP action for gravity is obtained from the Einstein-Hilbert
%%% action by replacing the Ricci scalar by Eq.~(\ref{ricscalar}) and 
%
\begin{equation}\label{a4}
S[e,A]=\int_\mathcal{M} |e|d^4x\, F^{IJ}_{\mu \nu} e^\mu_I e^\nu_J\,.
\end{equation}
This action is extremized under variations of the vierbein
$e^\mu_I$,  keeping $ A^I_{\mu J} $ fixed. Variation of the
determinant gives 
\begin{eqnarray}\label{delta_e}
\delta |e| = -|e|\,e^I_\mu \delta e^\mu_I.
\end{eqnarray}
Using the antisymmetry of $F^{IJ}_{\mu \nu} $, we can 
then derive the field equations quite easily,
\begin{equation}
2F^{IJ}_{\lambda\nu} e^\lambda_I - e^J_\nu F^{KL}_{\rho\sigma} e^\rho_K
e^\sigma_L = 0. 
\end{equation}
Contracting with $e_{\mu J}$, and using Eq.~(\ref{rictensor}), we
get the familiar form 
\begin{equation}%\label{a1.2.1}
R_{\mu \nu}-\frac{1}{2}g_{\mu \nu} R =0.
\label{einstein}
\end{equation} 

This equation would be the vacuum Einstein equation if we could show
that $ \nabla $ is torsion free, i.e., if $\Gamma^\alpha_{\mu \nu} $ 
is symmetric in $ \mu, \nu $. For
this purpose, we vary the action of Eq.~(\ref{a4}) again, but this
time with respect to the spin connection $ A^I_{\mu J}\,, $ keeping 
the vierbein fixed. To do this we first simplify the action using the
antisymmetry of the spin connection,
\begin{equation}
S[e,A] = \int |e|d^4x \left(2\partial_\mu A^{IJ}_\nu e^\mu_I e^\nu_J 
+ [A_\mu,A_\nu]^{IJ}e^\mu_I e^\nu_J\right).
\end{equation}
Variation with respect to $ A^{IJ}_\nu $ produces the equation
\begin{equation}\label{torsion_free}
-e_K^\alpha (\partial_\mu e^K_\alpha) e^\mu_I e^\nu_J 
- (\partial_\mu e^\mu_I) e^\nu_J - e^\mu_I (\partial_\mu e^\nu_J)
 + A^K_{\mu I} e^\mu_K e^\nu_J - A^K_{\mu I} e^\mu_J e^\nu_K = 0.
\end{equation}
In order to identify $ \Gamma $ in this equation, we contract
with  $ e^I_\rho e^J_\lambda $\,,
\begin{equation}
-e^\alpha_K (\partial_\rho e^K_\alpha) \delta^\nu_{\lambda} 
- (e^I_\rho \partial_\mu e^\mu_I) \delta^\nu_{\lambda} 
- (e^J_\lambda\partial_\rho e^\nu_J)
 + A^K_{\mu I} e^\mu_K e^I_\rho \delta^\nu_{\lambda} 
- A^K_{\lambda I} e^I_\rho e^\nu_K = 0\,.
\label{gamma.asym1}
\end{equation}
Antisymmetrising in $ \rho\,,\lambda $ and using the expression 
Eq.~(\ref{gammadef}) for $ \Gamma\,, $ we get
\begin{equation}
\delta^\nu_{\lambda}( \Gamma^\alpha_{\alpha \rho} 
- \Gamma^\alpha_{\rho\alpha} ) + \delta^\nu_{\rho}
( \Gamma^\alpha_{\lambda \alpha} 
- \Gamma^\alpha_{\alpha\lambda} )
 + \Gamma^\nu_{\rho\lambda}
 - \Gamma^\nu_{\lambda\rho} = 0\,,
\label{gamma.asym2}
\end{equation}
whose trace produces $ \Gamma^\alpha_{\alpha \rho} - 
\Gamma^\alpha_{\rho\alpha} = 0 $\,. Putting this back into 
Eq.~(\ref{gamma.asym2}) gives the desired equation,
\begin{equation}
\Gamma^\nu_{\rho\lambda} = \Gamma^\nu_{\lambda \rho}\,.
\end{equation}
Thus we can identify $\nabla$ as the unique metric-compatible 
torsion-free connection on the space-time. Although in usual General Relativity 
the torsion-free condition is imposed a priori, in the VEP formulation only 
metric-compatibility follows from the definition of $ \Gamma\,; $ 
the torsion-free condition comes from the equations of motion. Thus we
can now identify Eq.~(\ref{einstein}) with Einstein's equations in vacuum.

Once the connection has been identified as torsion-free, the spin connection 
becomes expressible as a function of tetrads, 
\begin{equation}
A^{IJ}_\mu =
\frac{1}{2}e^{\alpha I} e^{\beta J} 
e_{K[\alpha}\partial_\mu e^K_{\beta]} \,,
\label{spintetrad}
\end{equation}
using the definition of $ \Gamma $. 

So far we have considered the vacuum equations. In presence of matter, the total VEP action 
reads
\begin{equation}
S_{Total} = \frac{1}{16 \pi G} \int |e| d^4x F^{IJ}_{\mu\nu} e^\mu_I e^\nu_J 
+ S_M, 
\end{equation}
where $ S_M = \int |e| d^4x \mathcal{L}_M $ is the action for the matter.
%%% , with the Lagrangian density $ \mathcal{L}_M $ written in terms 
%%% of tetrads and the spin connection. 
The total VEP equation, obtained by variation with respect to 
the tetrad, is then
\begin{equation}
F^{IJ}_{\alpha \mu} e^\alpha_I - \frac{1}{2} e^J_\mu F^{KL}_{\alpha\beta} 
e^\alpha_K e^\beta_L = 8 \pi G T_{\mu\alpha}e^{\alpha J}\,,
\label{EE.tetrad}
\end{equation}
where $ T_{\mu \alpha} $ is the usual energy-momentum tensor for the matter. 
As before, we can contract this equation with the tetrad to obtain the familiar form, 
$ G_{\mu\nu} = 8 \pi G T_{\mu \nu} $. However, the inclusion of matter dictates 
whether the spin connection can be expressed entirely in terms of tetrads. More precisely, 
if the Lagrangian of the matter field contains the spin connection, we get 
a nonvanishing quantity on the right-hand side of Eq.~(\ref{torsion_free}) 
and thus the torsion-free condition is not obtained. This happens in particular
in case of fermionic matter, for which torsion remains an independent entity, 
as does the spin connection.
%%%%%%%%%%%%%%%%%%%%%%%%%%%%%%%%%
\section{VEP perturbations}\label{pert}
%%%%%%%%%%%%%%%%%%%%%%%%%%%%%%%%%
The spin connection is expressible in terms of the tetrad in the absence of 
matter, in particular fermionic matter, as Eq.~(\ref{spintetrad}) relies crucially 
on the connection being torsion-free, which in this formalism follows from 
%%% the equations of motion in vacuum. 
the matter action being independent of the spin connection $A^{IJ}_\mu\,.$ 
Thus we can relate the spin connection to the tetrad when the spin connection 
does not couple to matter. 

However,  perturbations of tetrad and spin connection around some background 
solution of the equations of motion should be considered as independent objects.
This is because perturbations are off-shell objects a priori and need not satisfy the
same equations as the background solutions, and even the background 
tetrad and spin connection cannot be related when there is a fermionic 
energy-momentum density on the background space-time.

We write the tetrad $ e^\mu_I $ as a sum of the background and perturbation,
\begin{equation}
e^\mu_I = \bar{e}^\mu_I + f^\mu_I,
\end{equation}
where $ f^\mu_I $ is much smaller than $ \bar{e}^\mu_I $ 
(more precisely, $ {\rm Tr}(\bar{e}^I_\mu f^\mu_I)\ll 1 $)\,. 
As before, we will denote background quantities by a bar on top. In 
order to calculate the co-tetrads, we use their definition 
$ e^\mu_I e^I_\nu = \delta^\mu_\nu = \bar{e}^\mu_I \bar{e}^I_\nu $, 
where the internal index is raised and lowered with $\eta_{IJ}$ as before. 
Thus we find 
\begin{equation}
e^I_\mu = \bar{e}^I_\mu - \bar{e}^J_\mu \bar{e}^I_\alpha f^\alpha_J.
\end{equation}
We will often denote $ - \bar{e}^J_\mu \bar{e}^I_\alpha f^\alpha_J $ as  
$ \widetilde{f}^I_\mu $. 
The space-time indices will be raised and lowered by the total 
space-time metric $ g_{\mu\nu} = e^I_\mu e^{\phantom I}_{I\nu}\,$ 
when needed. By writing the background metric as $ \bar{g}_{\mu\nu} 
= \bar{e}^I_\mu \bar{e}^{\phantom I}_{I\nu}\,, $ we can identify the 
metric perturbation $ h_{\mu\nu} $ in terms of the  background 
tetrad $\bar e_\mu^I$ and the tetrad perturbation $\widetilde{f}_\mu^I$ 
as
\begin{equation}\label{h_in_tetrads}
h_{\mu\nu} = \bar{e}_{I\mu} \widetilde{f}^I_{\nu} + 
\bar{e}_{I\nu} \widetilde{f}^I_{\mu}\,.
\end{equation}
The background now is any general space-time, not necessarily flat.
Let us also write the spin connection as a sum of its value 
 in the background space-time and a perturbation,
\begin{equation}
A^{IJ}_\mu = \bar{A}^{IJ}_\mu + a^{IJ}_\mu.
\end{equation}
However, since all components $\bar{A}^{IJ}_\mu$ of the background spin
connection may vanish, it is not sensible to treat $a^{IJ}_\mu$ as small 
perturbation. In particular, we will not neglect terms quadratic in $a^{IJ}_\mu$ 
when calculating the action. Thus we calculate the Christoffel 
symbols up to  first order in the  perturbation $ f $ as
\begin{equation}
\Gamma^\alpha_{\mu\nu} = \bar{\Gamma}^\alpha_{\mu\nu} 
+ \bar{e}^\alpha_I \partial_\mu \widetilde{f}^I_\nu + f^\alpha_I 
\partial_\mu \bar{e}^I_\nu + \bar{A}^{IJ}_\mu \bar{e}^J_\nu 
f^\alpha_I + a^{IJ}_\mu \bar{e}^J_\nu f^\alpha_I 
+ a^{IJ}_\mu \bar{e}^\alpha_I \widetilde{f}^J_\nu.
\end{equation}
Here $ \bar{\Gamma} $ corresponds to the background space-time,
\begin{equation}
\bar\Gamma^\alpha_{\mu\nu} = {\bar e}^\alpha_I \partial_\mu {\bar e}^I_\nu +
{\bar A}^I_{\mu J} {\bar e}^J_\nu {\bar e}^\alpha_I\,. 
\end{equation}

Let us also write the curvature in terms of $\bar A$ and $a\,,$ 
\begin{equation}
F^{IJ}_{\mu\nu} = \bar{F}^{IJ}_{\mu\nu} + \mathcal{F}^{IJ}_{\mu\nu}\,,
\end{equation}
where $ \bar{F} $ is the background curvature and $ \mathcal{F} $, the 
extra part due to perturbation, can be written in the form
\begin{equation}
\mathcal{F}^{IJ}_{\mu\nu} = \bar{D}_\mu a^{IJ}_\nu 
- \bar{D}_\nu a^{IJ}_\mu + [a_\mu,a_\nu]^{IJ}\,,
\end{equation}
with $ \bar{D} $ being the covariant derivative corresponding to the 
background spin connection $\bar A\,.$ The perturbed VEP action is thus 
\begin{equation}
S = \frac{1}{16 \pi G}\int |e|\, d^4x \left[F^{IJ}_{\mu\nu}
\bar{e}^\mu_I \bar{e}^\nu_J 
+ 2 F^{IJ}_{\mu\nu}\bar{e}^\mu_I f^\nu_J 
+ F^{IJ}_{\mu\nu}f^\mu_I f^\nu_J\right] + S_M \,.
\label{pert.action}
\end{equation}
The determinant $ |e| \equiv |\bar{e}^I_\mu + \widetilde f^I_\mu|\, $ 
is now a polynomial in $ f $\,.
The lowest order field equations will be 
of first order in $ f\,. $ The variation of the determinant produces
\begin{equation}
\delta |e| = -|e| (\bar{e}^K_\alpha + \widetilde{f}^K_\alpha)\delta f^\alpha_K\,,
\end{equation} 
using which we obtain field equations by varying the 
VEP action of Eq.~(\ref{pert.action}),
\begin{equation}
%\notag
(\bar{F}^{IJ}_{\alpha\mu} + \mathcal{F}^{IJ}_{\alpha\mu}) 
(\bar{e}^\alpha_I + f^\alpha_I) - \frac{1}{2} (\bar{e}^J_\mu + 
\widetilde{f}^J_{\mu}) 
( \bar{F}^{IJ}_{\alpha\beta} + \mathcal{F}^{IJ}_{\alpha\beta}) 
(\bar e^\alpha_K + f^\alpha_K) (\bar e^\beta_L + f^\beta_L) = 
8\pi G T_{\mu\nu} e^{\nu J}\,.
\end{equation}
Of course, we could have obtained these directly from Eq.~(\ref{EE.tetrad}) 
by replacing $e \to \bar e+\widetilde f\,$ and $A \to \bar A+a\,.$ However,  
it is useful to construct the action for the perturbations, as we will later consider
matter couplings and a tetrad version of Fierz-Pauli equations.

Subtracting the VEP equation for the background, we get the equation of 
motion for the VEP perturbations,
\begin{align}\label{vep_pert_matter}
\notag
\bar{F}^{IJ}_{\alpha\mu} f^\alpha_I - \frac{1}{2} 
\bar{F}^{KL}_{\alpha\beta} \left[ \widetilde{f}^J_\mu 
\bar{e}^\alpha_K \bar{e}^\beta_L + 2 \bar{e}^J_\mu f^\alpha_K \bar{e}^\beta_L \right] 
+\mathcal{F}^{IJ}_{\alpha\mu} \left( \bar{e}^\alpha_I + f^\alpha_I \right) & \\
- \frac{1}{2} \mathcal{F}^{KL}_{\alpha\beta} \left[\bar{e}^J_\mu 
\bar{e}^\alpha_K \bar{e}^\beta_L + \widetilde{f}^J_\mu 
\bar{e}^\alpha_K \bar{e}^\beta_L + 2 \bar{e}^J_\mu f^\alpha_K 
\bar{e}^\beta_L \right] &= 8 \pi G \left(T_{\mu\nu} e^{\nu J} - 
\bar{T}_{\mu\nu}\bar{e}^{\nu J}\right)\,.
\end{align}

This is a generic equation in the sense that we have not considered any particular 
background, or required that the background be flat, so this is the equation of 
perturbations around a general background space-time. In order to compare this 
equation with the metric perturbation equations of Eq.~(\ref{metric_pert_matter.2}), 
let us consider some specific types of matter field theories.  Whether or not 
the perturbation of the spin connection depends on tetrads relies entirely 
on the type of field considered. For non-fermionic matter, 
we will be able to use Eq.~(\ref{spintetrad}) for the spin connection and thus
write Eq.~(\ref{vep_pert_matter}) in terms of the tetrad and its perturbation. 
Let us investigate the VEP equation with scalar, electromagnetic and 
fermionic fields as examples of background matter. 

%%%%%%%%%%%%%%%%%%%%%%
\section{VEP equation with matter fields}\label{matter}\label{fields}
%%%%%%%%%%%%%%%%%%
%%%%%%%%%%%%%%%%%%%%%
\subsection{Scalar field}
%%%%%%%%%%%%%%%%%%%%%
The action for a massive real scalar field $\phi$ is 
\begin{equation}
S_M =\int |e| d^4x \left[ -\frac12 \nabla_\mu \phi \nabla^\mu \phi - \frac12 m^2 \phi^2 \right].
\end{equation}
The equation for the background vierbein with this matter field is given by
\begin{eqnarray}
\bar{F}^{IJ}_{\alpha \mu} \bar{e}^\alpha_I - \frac12 \bar{e}^J_\mu 
\bar{F}^{KL}_{\alpha \beta} \bar{e}^\alpha_K \bar{e}^\beta_L =
8 \pi G \left[ \nabla_\mu \phi \nabla_\nu \phi - \frac12 \bar{e}^I_\mu 
\bar{e}_{\nu I} \left( \bar{e}^\alpha_J \bar{e}^{\beta J} 
\nabla_\alpha \phi \nabla_\beta \phi + m^2 \phi^2 \right) \right] \bar{e}^{\nu J}\,.
\label{vepeq.scalar}
\end{eqnarray}
The right-hand side of this equation contains the background 
energy-momentum tensor of the field. Taking the trace of the 
equation by contraction with $ \bar{e}^\mu_J\,, $ and then eliminating 
the trace, the equation can be written as
\begin{equation}\label{riemann_scalar}
\bar{F}^{IJ}_{\alpha \mu} \bar{e}^\alpha_I = 8 \pi G 
\left[ \nabla_\mu \phi \nabla_\alpha \phi \bar{e}^{\alpha J} 
+ \frac12 m^2 \bar{e}^J_{\mu} \phi^2 \right]\,. 
\end{equation}
For the massive scalar field, the VEP equation for vierbein perturbation, 
Eq.~ (\ref{vep_pert_matter}), can be written as
\begin{align}
\notag
\bar{F}^{IJ}_{\alpha\mu} f^\alpha_I &- \frac{1}{2} \bar{F}^{KL}_{\alpha\beta} 
\left[ \widetilde{f}^J_\mu \bar{e}^\alpha_K \bar{e}^\beta_L 
+ 2 \bar{e}^J_\mu f^\alpha_K \bar{e}^\beta_L \right] + \mathcal{F}^{IJ}_{\alpha\mu} 
\left( \bar{e}^\alpha_I + f^\alpha_I \right) \\
\notag
&- \frac{1}{2} \mathcal{F}^{KL}_{\alpha\beta} 
\left[\bar{e}^J_\mu \bar{e}^\alpha_K \bar{e}^\beta_L 
+ \widetilde{f}^J_\mu \bar{e}^\alpha_K \bar{e}^\beta_L 
+ 2 \bar{e}^J_\mu f^\alpha_K \bar{e}^\beta_L \right]\\
&\qquad = -4 \pi G \left[ \left( \widetilde{f}^J_\mu \bar{e}^\alpha_K 
\bar{e}^{\beta K} + 2 \bar{e}^J_\mu \bar{e}^\alpha_K 
f^{\beta K} \right)\nabla_\alpha \phi \nabla_\beta \phi  
+\widetilde{f}^J_\mu m^2\phi^2 \right].
\end{align}

Interestingly, for the scalar field, the matter part can be completely removed
from this equation. We use Eq.~(\ref{riemann_scalar}) to write the equation in the form 
\begin{eqnarray}\label{vep_pert_scalar_1}
\bar{F}^{IJ}_{\alpha\mu} f^\alpha_I + \mathcal{F}^{IJ}_{\alpha\mu} 
\left( \bar{e}^\alpha_I + f^\alpha_I \right) - \frac{1}{2} 
\mathcal{F}^{KL}_{\alpha\beta} \left[\bar{e}^J_\mu 
\bar{e}^\alpha_K \bar{e}^\beta_L + \widetilde{f}^J_\mu 
\bar{e}^\alpha_K \bar{e}^\beta_L + 2 \bar{e}^J_\mu f^\alpha_K 
\bar{e}^\beta_L \right] = 0.
\end{eqnarray}
Also, since the Lagrangian of scalar field does not contain the spin connection, 
variation of the action with respect to the connection yields the torsion-free condition
as in vacuum. This enables us to express both the background spin connection 
and its perturbation in terms of $ e $ and $ f $. At the lowest order, the 
perturbation $ a^{IJ}_\mu $ turns out to be of first order in $ f $ and 
is given by
\begin{equation}\label{a in e0f}
a^{IJ}_\mu = \frac{1}{2}\left(\bar{e}^{\alpha I} \bar{e}^{\beta J} 
- \bar{e}^{\beta I} \bar{e}^{\alpha J}\right)
\left[ \partial_\mu \left(\bar{e}^K_\beta \widetilde{f}_{\alpha K}\right)+
\partial_\beta\left(\bar{e}^K_\mu \widetilde{f}_{\alpha K}\right)- 
\partial_\alpha\left(\bar{e}^K_\beta \widetilde{f}_{\mu K}\right)\right].
\end{equation}
Hence $ \mathcal{F} $ is also of first order in $ f $ at the lowest order. 
Thus neglecting higher order terms, Eq.~(\ref{vep_pert_scalar_1}) can be 
rewritten as
\begin{eqnarray}\label{vep_pert_scalar}
\bar{F}^{IJ}_{\alpha\mu} f^\alpha_I + \mathcal{F}^{IJ}_{\alpha\mu} 
\bar{e}^\alpha_I - \frac{1}{2} \mathcal{F}^{KL}_{\alpha\beta} 
\bar{e}^J_\mu \bar{e}^\alpha_K \bar{e}^\beta_L = 0.
\end{eqnarray}

%%%%%%%%%%%%%%%%%%%%%%%%%%%%%%%%%%%%%% ELECTROMAGNETIC %%%%%%%%%%%%%%%%%%%%%%%%%%%%%%%%%%%%%%%%%
%%%%%%%%%%%%%%%%%%%%%
\subsection{Electromagnetic field}
%%%%%%%%%%%%%%%%%%%%%
Next we consider the VEP equation with electromagnetic energy-momentum as source. 
Let us denote the electromagnetic field strength as $\mathbb{F}_{\mu\nu}\,,$ then
the Lagrangian of the field can be written as $ -\frac14 \mathbb{F}_{\mu\nu} 
\mathbb{F}^{\mu\nu} $. The energy-momentum tensor for this is 
\begin{equation}
T_{\mu\nu} = \mathbb{F}_{\mu\alpha}\mathbb{F}^{~ \alpha}_\nu - \frac14 e^I_\mu e_{\nu I} \mathbb{F}_{\alpha\beta} \mathbb{F}^{\alpha\beta}\,,
\end{equation}
which is traceless and independent of the connection components. Thus 
the variation with respect to the connection again leads to the torsion-free 
condition as in the case of the scalar field. We can then 
%%% eliminate the trace of the  VEP equation and also neglect 
%%% higher order terms in $ f $.
write Eq.~(\ref{vep_pert_matter}) with the electromagnetic field as matter source,
\begin{equation}
\bar{F}^{IJ}_{\alpha\mu} f^\alpha_I + \mathcal{F}^{IJ}_{\alpha\mu} 
\bar{e}^\alpha_I = 8\pi G \left[ \mathbb{F}_{\mu\lambda}\mathbb{F}_{\nu \rho} 
\left( \bar{e}^\lambda_I \bar{e}^{\nu J} f^{\rho I} + \bar{e}^\rho_I \bar{e}^{\nu J} 
f^{\lambda I} \right) - \frac14 \tilde{f}^J_\mu \mathbb{F}_{\alpha \beta} 
\mathbb{F}^{\alpha \beta} - \mathbb{F}_{\alpha \beta} 
\mathbb{F}^\alpha_{~\sigma} \bar{e}^J_\mu \bar{e}^\beta_K f^{\sigma K} \right]\,.
\end{equation}
Unlike for the scalar field, here we cannot eliminate the matter part. Thus 
the above equation is the final form of the VEP equation under perturbation 
with electromagnetic field as source.

%%%%%%%%%%%%%%%% FERMION %%%%%
\subsection{Fermionic field}
%%%%%%%%%%%%%%%%%%
The action for a  Dirac field $ \psi $ on a curved background  
is given by~\cite{Fock:1929vt, Pollock:2010zz,Hehl:1976kj}
\begin{equation}
S_F =  \int |e| d^4x \left[ i \bar{\psi}\gamma^K e^\mu_K \left(\partial_\mu 
+ \frac{i}{4} A^{IJ}_\mu \sigma_{IJ}\right)\psi - m\bar{\psi}\psi \right]\,,
\end{equation}
where $ \sigma_{IJ} = \frac i2 [\gamma_I,\gamma_J]$\,. 
The perturbation equation Eq.~(\ref{vep_pert_matter}) can be written as
\begin{align}
\notag
\bar{F}^{IJ}_{\alpha\mu} f^\alpha_I - \frac{1}{2} 
\bar{F}^{KL}_{\alpha\beta} \left[ \widetilde{f}^J_\mu 
\bar{e}^\alpha_K \bar{e}^\beta_L + 2 \bar{e}^J_\mu f^\alpha_K 
\bar{e}^\beta_L \right] +\mathcal{F}^{IJ}_{\alpha\mu} 
\left( \bar{e}^\alpha_I + f^\alpha_I \right) &\\
- \frac{1}{2} \mathcal{F}^{KL}_{\alpha\beta} 
\left[\bar{e}^J_\mu \bar{e}^\alpha_K \bar{e}^\beta_L 
+ \widetilde{f}^J_\mu \bar{e}^\alpha_K \bar{e}^\beta_L 
+ 2 \bar{e}^J_\mu f^\alpha_K \bar{e}^\beta_L \right] &= 
8 \pi G \left(T_{\mu\nu}e^{\nu J} - \bar{T}_{\mu\nu} 
\bar{e}^{\nu J} \right)\,.
\end{align}
The right-hand side of this equation works out to be 
\begin{align}
T_{\mu\nu}e^{\nu J} - \bar{T}_{\mu\nu} \bar{e}^{\nu J} 
=& \frac 14 \bar{\psi} \gamma^J a^{KL}_\mu \sigma_{KL} \psi 
+ \frac14 \bar{e}^J_\mu \bar{e}^\nu_K \bar{\psi} 
\gamma^K a^{LM}_\nu \sigma_{LM} \psi
\notag \\
\notag
& - \bar{e}^J_\mu f^\nu_K \left( i\bar{\psi}\gamma^K 
\left(\partial_\nu + \frac i4\left( \bar{A}^{LM}_\nu + a^{LM}_\nu \right)
\sigma_{LM}\right)\psi \right)\\
& - \bar{e}^\nu_K \widetilde{f}^J_\mu \left( i\bar{\psi}\gamma^K
\left(\partial_\nu + \frac i4 \left( \bar{A}^{LM}_\nu + a^{LM}_\nu \right)
\sigma_{LM} \right)\psi \right) 
+ m \widetilde{f}^J_\mu \bar{\psi}\psi\,.
\label{Dirac.pert_matter}
\end{align}
Next we vary the action with respect to $a_\mu^{IJ}$\,. The resulting
equation is 
\begin{equation}
%\notag
-e_K^\alpha (\partial_\mu e^K_\alpha) e^\mu_I e^\nu_J 
- (\partial_\mu e^\mu_I) e^\nu_J - e^\mu_I (\partial_\mu e^\nu_J)
 - A^K_{\mu I} e^\mu_K e^\nu_J + A^K_{\mu I} e^\mu_J e^\nu_K 
%&\\
 -  \pi G\,\bar{\psi} \gamma^K e^\nu_K \sigma_{IJ}\psi = 0\,,
\label{}
\end{equation}
where we have written the equation in terms of the total vierbein 
and an overall factor of two has been eliminated. 
It is easy to show from this equation that the connection is not torsion 
free. If we contract  with $ e^I_\rho e^J_\lambda $ and then antisymmetrize 
in $ \rho,\lambda $ as in Eq.~(\ref{gamma.asym2}), we get
\begin{equation}\label{gamma_fermion}
\delta^\nu_{~ \lambda}\left(\Gamma^\alpha_{\rho\alpha} 
- \Gamma^\alpha_{\alpha\rho}\right) + \delta^\nu_{~\rho}
\left( \Gamma^\alpha_{\alpha\lambda} - \Gamma^\alpha_{\lambda\alpha}\right) 
+ \left( \Gamma^\nu_{\lambda\rho} - \Gamma^{\nu}_{\rho\lambda} \right)  
= 2 \pi G~\bar{\psi} \gamma^K e^\nu_K \sigma_{IJ} e^I_\lambda e^J_\rho \psi\,.
\end{equation}
Trace of this equation produces
\begin{equation}
\Gamma^\alpha_{\rho\alpha} - \Gamma^\alpha_{\alpha\rho} 
= -3i\pi G \bar{\psi}\gamma_I e^I_\rho \psi,
\end{equation}
where we have used $ \gamma^K \sigma_{IK} = -3i\gamma_I\,.$ 
Using this in Eq.~(\ref{gamma_fermion}) we find
\begin{equation}
\Gamma^\nu_{\lambda\rho} - \Gamma^{\nu}_{\rho\lambda} 
= 2 \pi G~\bar{\psi} \gamma^K e^\nu_K \sigma_{IJ} 
e^I_\lambda e^J_\rho \psi + 3i\pi G \bar{\psi}
\gamma_I \left( \delta^\nu_\lambda e^I_\rho - 
\delta^\nu_\rho e^I_\lambda \right) \psi \,.
\label{Dirac.torsion}
\end{equation}
This equation implies that $ \Gamma $ is not torsion-free and the source 
of torsion is clearly the fermionic field as the equation indicates. 
It is easy to see that this 
expression for torsion can be written in the irreducible form as 
in~\cite{Nikiforova:2009qr,Nair:2008yh}, with all the symmetries and 
trace identities.

If we have non-fermionic matter on a flat background, the VEP perturbation 
equations can be identified with the linearized Einstein equations, as we 
will see in the next section.

%%%%%%%%%%%%%%%%% EIERZ-PAULI%%%%%%%%%%%%
\section{Fierz-Pauli equation}\label{mass}
%%%%%%%%%%%%%%%%%%%%%%%%%%
Massive spin-2 fields obey the Fierz-Pauli equation~\cite{Fierz:1939ix}.
In this section we will write the Fierz-Pauli equation and the 
corresponding action in the VEP formalism. An action for massive 
gravitons, from which the Fierz-Pauli equation may be derived, can be
written as
\begin{equation}
S = \frac{1}{16 \pi G}\int d^4x \left[ -\frac{1}{2}\partial_\lambda h_{\mu\nu} 
\partial^\lambda h^{\mu\nu} + \partial_\mu h_{\nu\lambda} 
\partial^\nu h^{\mu\lambda} %&
- \partial_\mu h_{\mu\nu} 
\partial_\nu h + \frac{1}{2}\partial_\lambda h \partial^\lambda h
%\nonumber\\
%&
 -\frac{1}{2}m^2(h_{\mu\nu}h^{\mu\nu} - h^2) \right]\,,
\label{FP.action}
\end{equation}
We can obtain the Fierz-Pauli equation for massive gravity by varying 
$h_{\mu\nu}\,,$
\begin{equation}\label{fp} 
\square h_{\mu\nu} - \partial_\sigma\partial_\nu h^\sigma_{~\mu} 
- \partial_\sigma\partial_\mu h^\sigma_{~\nu} + 
\eta_{\mu\nu}\partial_\rho \partial_\sigma h^{\rho\sigma} 
+ \partial_\mu\partial_\nu h - \eta_{\mu\nu} \square h 
- m^2(h_{\mu\nu} - \eta_{\mu\nu}h) = 0\,.
\end{equation}
Usually this equation is written on a flat background space-time, and 
more specifically, with a background Minkowski metric, as we have 
done here. In order to get a theory of massive VEP perturbations, we 
add the Fierz-Pauli mass term to the action in terms of the VEP variables. 
Let us write the mass term for a background Minkowski metric as
\begin{equation}
%\nonumber
-\frac{1}{2}m^2(h_{\mu\nu} h^{\mu\nu} - h^2) =
-m^2 \left( \bar{e}^J_\mu \bar{e}^I_\nu f^\nu_J f^\mu_I 
+ \bar{e}_{\mu I} \bar{e}^I_\nu f^\nu_J f^{\mu J} 
- 2 \bar{e}^I_\mu \bar{e}^J_\nu f^\mu_I f^\nu_J \right)\,,
\label{mass.metric}
\end{equation}
We note that this term depends 
only on the tetrad perturbations, not on the spin connection.

In the absence of matter, the connection is torsion-free, and 
thus $ \mathcal{F} $ is of first order in $ f $ at the lowest order. 
Therefore Eq.~(\ref{vep_pert_matter}) reduces, at the lowest order in $f$\,,
to 
\begin{equation}
\mathcal{F}^{IJ}_{\alpha \mu} \bar{e}^\alpha_I 
- \frac12 \bar{e}^J_\mu \mathcal{F}^{KL}_{\alpha\beta} 
\bar{e}^\alpha_K \bar{e}^\beta_L = 0\,
\end{equation}
in the absence of matter on a flat background, i.e. 
$\bar F = 0\,.$
We can derive this equation from the action
\begin{equation}
S_{eff} = \frac{1}{16 \pi G}\int |e| d^4x 
\left( 2 \mathcal{F}^{IJ}_{\alpha \mu}\bar{e}^\alpha_I f^\mu_J 
- \bar{e}^J_\mu f^\mu_J \mathcal{F}^{KL}_{\alpha \beta}
\bar{e}^\alpha_I \bar{e}^\beta_L \right).
\end{equation}
The massive VEP action on a flat, matter-free background, is thus
\begin{equation}
S = \frac{1}{16 \pi G} \int |e| d^4x \left[ 2 \mathcal{F}^{IJ}_{\alpha \mu}
\bar{e}^\alpha_I f^\mu_J - \bar{e}^J_\mu f^\mu_J 
\mathcal{F}^{KL}_{\alpha \beta}\bar{e}^\alpha_I \bar{e}^\beta_L - m^2 
\left( \bar{e}^J_\mu \bar{e}^I_\nu f^\nu_J f^\mu_I 
+ \bar{e}_{\mu I} \bar{e}^I_\nu f^\nu_J f^{\mu J} 
- 2 \bar{e}^I_\mu \bar{e}^J_\nu f^\mu_I f^\nu_J \right) \right]\,,
\end{equation}
which leads to the massive VEP equation by varying $f\,,$
\begin{equation}
\mathcal{F}^{IJ}_{\alpha \mu} \bar{e}^\alpha_I 
- \frac12 \bar{e}^J_\mu \mathcal{F}^{KL}_{\alpha\beta} 
\bar{e}^\alpha_K \bar{e}^\beta_L - m^2 \left( \bar{e}^I_\mu 
\bar{e}^J_\nu f^\nu_I + \bar{e}_{\mu I} \bar{e}^I_\nu f^{\nu J} 
- 2 \bar{e}^J_\mu \bar{e}^I_\nu f^\nu_I \right) = 0\,.
\end{equation}
As we have already mentioned, the Fierz-Pauli mass term is usually
added to the linearized equations for perturbations --- on a flat background,
with the Minkowski metric --- without any fermionic matter. In 
that case, the spin connection can be expressed as a function of the tetrads,
as we have seen. Thus we could as well derive the corresponding
equations for vierbein perturbations simply by replacing the metric and
its perturbation in terms of those of the tetrads. 

However, the way we have written the Fierz-Pauli equation and action allows
an easy generalization to arbitrary curved backgrounds.
We add the mass term on the right-hand side of Eq.~(\ref{mass.metric}) 
to the generalized VEP action Eq.~(\ref{pert.action}) to get
\begin{align}
S = \frac{1}{16 \pi G} \int |e|\, &d^4x \left[F^{IJ}_{\mu\nu}\bar{e}^\mu_I \bar{e}^\nu_J 
+ 2 F^{IJ}_{\mu\nu}\bar{e}^\mu_I f^\nu_J 
+ F^{IJ}_{\mu\nu}f^\mu_I f^\nu_J \right.\, \notag\\
& \qquad -m^2 \left( \bar{e}^J_\mu \bar{e}^I_\nu 
\left.f^\nu_J f^\mu_I + \bar{e}_{\mu I} \bar{e}^I_\nu f^\nu_J f^{\mu J} 
- 2 \bar{e}^I_\mu \bar{e}^J_\nu f^\mu_I f^\nu_J \right)\right] \,.
\label{pert.action.mass}
\end{align}
This is the VEP form of the Fierz-Pauli action in terms of a single set of tetrad
fields, written as background plus perturbation. 
Written in this form, the action is not restricted to a flat background space-time,
and we can add any type of matter source, including fermionic matter, to this action. 

While this paper was in circulation as a preprint, a couple of papers have appeared 
regarding perturbations of bimetric gravity on an arbitrary background 
spacetime~\cite{Bernard:2015uic, Cusin:2015tmf}. Since these papers do
not consider a vierbein formulation of the theory, their results cannot be easily
compared with ours. However, one expects that their results and ours can be related
by a careful comparison, which we leave for a future work.

%%%%%%%%%%%%%%%%%%%%%%%%%%%%%%%%%%%%%%%%%%%%%%%%%%%%%%

\end{document}